\documentclass[a4paper]{jpconf}
\usepackage{amssymb,graphicx,epsfig,color}
\usepackage{amsmath,bbm,amsfonts}
\def\({\left(} \def\){\right)} \def\[{\left[} \def\]{\right]}
  
\def\Tr{\hbox{Tr}}
\begin{document}
\title{Information/disturbance trade-off in single and sequential 
measurements on a qudit signal}
\author{Marco G Genoni and Matteo G A Paris}
\address{Dipartimento di Fisica, Universit\'a degli studi di Milano, Italia}
\ead{marco.genoni@studenti.unimi.it, matteo.paris@fisica.unimi.it}
\begin{abstract}
We address the trade-off between information gain and state disturbance 
in measurement performed on qudit systems and devise a class of optimal 
measurement schemes that saturate the ultimate bound imposed by quantum 
mechanics to estimation and transmission fidelities. The schemes are 
{\em minimal}, {\em i.e.} they involve a single additional probe qudit, 
and {\em optimal}, {\em i.e.} they provide the maximum amount of information 
compatible with a given level of disturbance. The performances of optimal 
single-user schemes in extracting information by sequential measurements 
in a $N$-user transmission line are also investigated, and the optimality 
is analyzed by explicit evaluation of fidelities. We found that the estimation 
fidelity does not depend on the number of users, neither for single-measure 
inference nor for collective one, whereas the transmission fidelity 
decreases with $N$. The resulting trade-off is no longer optimal and 
degrades with increasing $N$. We found that optimality can be restored by 
an effective preparation of the probe states and present explicitly 
calculations for the 2-user case.
\end{abstract}
\section{Introduction}
Any measurement aimed to extract information about a quantum state alters 
the state itself, {\em i.e.} introduces a disturbance \cite{hh0}. In
addition, quantum information cannot be perfectly copied, neither locally 
\cite{nocl} nor at distance \cite{telecl}. Overall, there is an 
information/disturbance trade-off which unavoidably limits the accuracy
of any kind of measurement, independently of the implementation scheme \cite{KB}.  
On the other hand, in a multiuser transmission line each user should 
decode the transmitted symbol and leave the carrier for the subsequent user.  
Indeed, what they need is a device able to retrieve as much 
information as possible, without destroying the carrier.
Since in a quantum channel symbols are necessarily encoded in states of
a physical system, the ultimate bounds on the channel performances are
posed by quantum mechanics. 
\par
The trade-off between information gain and quantum state disturbance can
be quantified using fidelities. Let us describe a generic scheme for
indirect measurement as a quantum operation, {\em i.e.} without
referring to any explicit unitary realization.  The operation is
described by a set of {\em measurement operators $\{A_k\}$}, with the
trace-preserving 
condition $\sum_k A^\dag_k A_k= {\mathbbm I}$. The probability-operator
measure (POVM) of the measurement is given by $\{\Pi_k\equiv A_k^\dag
A_k\}$, whereas its action on the input state is expressed as $\varrho
\rightarrow \sum_k A_k\varrho A_k^\dag$. This means that, if $\varrho$
is the initial quantum state of the system under investigation, the
probability distribution of the outcomes is given by
$p_k=\hbox{Tr}[\varrho\: \Pi_k]= \hbox{Tr}[\varrho\: A^\dag_k A_k]$. 
The conditional output state, after having detected the outcome
$k$, is given by $\sigma_k= \: A_k\varrho A_k^\dag/p_k$, whereas 
the overall quantum state after the measurement is described by the
density matrix $\sigma=\sum_k p_k \: \sigma_k=\sum_k A_k\varrho
A_k^\dag$.  
\par
Suppose you have a quantum system prepared in a pure state
$|\psi\rangle$. If the outcome $k$ is observed at the output of the
measuring device, then the estimated signal state is given by
$|\phi_k\rangle$ (the typical inference rule being $k \rightarrow
|\phi_k\rangle$ with $|\phi_k\rangle$ given by the set of eigenstates of
the measured observable), whereas the conditional state $|\psi_k\rangle
= 1/\sqrt{p_k} A_k |\psi\rangle$ is left for the subsequent user. The
amount of disturbance is quantified by evaluating the overlap of the
conditional state $|\psi_k\rangle$ to the initial one $|\psi\rangle$,
whereas the amount of information extracted by the measurement
corresponds to the overlap of the inferred state $|\phi_k\rangle$ to the
initial one.  The corresponding fidelities, for a given input signal
$|\psi\rangle$, are given by
\begin{align}
F_\psi &= \sum_k p_k \frac{|\langle\psi|A_k |\psi\rangle |^2}{p_k} 
= \sum_k |\langle \psi|A_k|\psi\rangle |^2 \label{F_psi}
\\ G_\psi & = \sum_k p_k |\langle\psi|\phi_k\rangle |^2  
\label{G_psi}\;,
\end{align}
where we have already performed the average over the outcomes. The
relevant quantities to assess the performances of the device are given
by the average fidelities
\begin{align}
F = \int_{\mathbbm A} d\psi \: F_\psi \qquad
G = \int_{\mathbbm A} d\psi \: G_\psi
\label{fids}\;,
\end{align}
which are obtained by averaging $F_\psi$ and $G_\psi$ over the possible
input states, {\em i.e.} over the alphabet ${\mathbbm A}$ of
transmittable symbols (states). $F$ will be referred to as the {\em
transmission fidelity} and $G$ to as the {\em estimation fidelity}.
\par
Let us first consider two extreme cases.  If nothing is done, the signal
is preserved and thus $F=1$. However, at the same time, our estimation
has to be random and thus $G=1/d$ where $d$ is the dimension of the
Hilbert space. This corresponds to a {\em blind} quantum repeater
\cite{szeged} which re-prepares any quantum state received at the input,
without gaining any information on it.  The opposite case is when the
maximum information is gained on the signal, {\em i.e.} when the optimal
estimation strategy  for a single copy is adopted
\cite{popescu,acin,bruss}.  In this case $G=2/(d+1)$, but then the
signal after this operation cannot provide any more information on the
initial state and thus $F=2/(d+1)$. 
In between these two extrema there are intermediate cases, {\em i.e.}
quantum measurements providing only partial information while partially
preserving the quantum state of the signal for subsequent users.  
\par
The fidelities $F$ and $G$ are not independent on each other. 
Assuming that ${\mathbbm A}$ corresponds to the set of 
{\em all} possible quantum states, Banaszek \cite{KB} has 
explicitly evaluated the expressions of fidelities in terms of 
the measurement operators, thus rewriting Eqs. (\ref{fids}) as 
\begin{align}
F &=  \frac{1}{d(d+1)} \left(d+ \sum_k \left| \hbox{Tr} \left[A_k\right]
\right|^2\right)  
\label{fidseF}\:, \\ 
G &= \frac{1}{d(d+1)} \left(d+ \sum_k \langle \phi_k | \Pi_k | 
\phi_k\rangle \right)
\label{fidseG}\;,
\end{align}
where $|\phi_k\rangle$ is the set of states used to estimate the 
initial signal. Of course, the estimation fidelity is maximized 
upon choosing $|\phi_k\rangle$ as the eigenvector of the POVM-element 
$\Pi_k$  corresponding to the maximum eigenvalues.
\par
Using Eqs. (\ref{fidseF}) and (\ref{fidseG})it is possible to derive 
the ultimate bound that fidelities should satisfy according to quantum mechanics. 
For randomly distributed $d$-dimensional signals, {\em i.e.} 
when the alphabet ${\mathbbm A}$ corresponds to the set of 
{\em all} quantum states of a qudit, the information-disturbance 
trade-off reads as follows \cite{KB}
\begin{align}
(F - F_0)^2 + d^2 ( G - G_0)^2 
+ 2 (d-2)(F - F_0)( G - G_0)
\leq \frac{d-1}{(d+1)^2} \label{Dbound}\;,
\end{align}
where $F_0=\frac12(d+2)/(d+1)$ and $G_0=\frac12\:3/(d+1)$.
From Eq. (\ref{Dbound}) one may derive the maximum transmission fidelity
compatible with a given value of the estimation fidelity or, in other
words, the minimum unavoidable amount of noise that is added, in
average, to a set of random qudits if one wants to acquire a given
amount of information.
Notice that the trade-off crucially depends on the alphabet of
transmittable symbols. Ultimate bound on fidelities have been derived
for different set of signals. These include many copies of identically
prepared pure qubits \cite{KB2}, a single copy of a pure state generated
by independent phase-shifts \cite{MistaPRA72}, an unknown spin coherent
state \cite{msqph}, and single copy of an unknown maximally entangled
state \cite{MS06}.  Optimal measurement schemes, which saturate the
bounds, have been also devised \cite{MGenoni,MistaPRA05,sbqph} and
implemented \cite{sc06}.
\par
In this paper we review our results \cite{MGenoni} on the unitary realizations
of optimal estimation of qudits and present novel results about the
information/disturbance trade-off in a multiuser transmission line. As we will
see, our schemes are {\em minimal}, since they involve a single additional
probe qudit, and {\em optimal} {\em i.e.} the corresponding fidelities
saturate the bound (\ref{Dbound}). We investigate their performances in
extracting information  in a multiuser transmission line where the unknown
signal state $|\psi\rangle$ is measured sequentially by $N$ users, each of
them using optimal single-user device.
\par
The paper is structured as follows. In Section \ref{s:Opt} the optimal
scheme for the estimation of a generic qudit is reviewed in details. In
Section \ref{s:Seq} we analyze the trade-off for a multiuser
transmission line, focusing attention on low-dimensional qudit
($d=\{2,3,4\}$). The optimal trade-off for qubit in a $2$-user
transmission line is explicitly evaluated. Section \ref{s:outro} closes
the paper with some concluding remarks.
\section{Minimal implementation of optimal measurement schemes for qudit}\label{s:Opt}
\begin{figure}[h!]
\centerline{\includegraphics[width=0.45\textwidth]{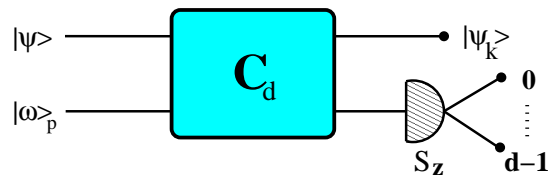}}
\caption{Minimal implementation of optimal measurement scheme for qudits.} 
\label{f:schd}
\end{figure}

In this section we describe a class of measurement schemes devised to estimate 
the state of a random qudit without its destruction \cite{MGenoni}. The
schemes are minimal, since they involve a single additional probe qudit, and
optimal because they saturate the bound (\ref{Dbound}).  The
measurement scheme is shown in Fig. \ref{f:schd}.  The signal qudit is
coupled with a probe qudit prepared in the state 
\begin{align}
|\omega\rangle_p = \cos\theta |0\rangle_p + \gamma \sin\theta \: 
\frac{1}{\sqrt{d}} \sum_{s=0}^{d-1} | s\rangle_p
\label{omd}
\end{align}
where 
$ \gamma = \left(\sqrt{1+d \tan^2\theta}-1\right)/ \sqrt{d} \tan\theta $
is a normalization factor. The interaction is given by the unitary gate
${\mathbf C}_d$ that acts as ${\mathbf C}_d|i\rangle |s\rangle_p =
|i\rangle |i\oplus s\rangle_p$ where $\oplus$ denotes sum modulo $d$
\cite{delg}.  After the interaction the spin of the probe is measured in
the $z$-direction.
\par\noindent
The measurement operators are 
$A_k = {}_p \langle k | {\mathbf C}_d | \omega \rangle_p = \sum_{ij} 
(A_k)_{ij} |i\rangle\langle j|$
where 
\begin{eqnarray}
(A_k)_{ij} = \delta_{ij} \left[ \delta_{kj} \cos\theta + \frac{\gamma}{\sqrt{d}}\sin\theta \right]
\label{akd_ij}\;.
\end{eqnarray}
The inference rule is $k \rightarrow |k\rangle$, where $|k\rangle$ is an 
eigenstate of $S_z$.
The fidelities are evaluated using 
Eqs. (\ref{fidseF}), (\ref{fidseG}) and (\ref{akd_ij}), arriving at
\begin{eqnarray}
F&=& \frac{1}{d+1} \left[1 + \left(\cos\theta + \gamma \sqrt{d} 
\sin\theta \right)^2\right] 
\label{DfidsF}\;, \\
G&=& \frac{1}{d+1} \left[1 + \left(\cos\theta + \frac{\gamma}{\sqrt{d}} 
\sin\theta \right)^2\right]
\label{DfidsG}\;.
\end{eqnarray}
Upon inserting Eqs. (\ref{DfidsF}) and (\ref{DfidsG}) into Eq. (\ref{Dbound})
we found that the bound is saturated. In
other words, the scheme of Fig. \ref{f:schd} provides an optimal measurement
scheme for qudits.
\par 
The optimal preparation of the probe state can be intuitively understood as
follows: for $\theta=0$ the elements of the measurement operators reduce to
$(A_k)_{ij}=\delta_{ij} \delta_{kj}$, which lead to fidelities $F=2/(d+1)$ and
$G=2/(d+1)$, {\em i.e.} to the extreme case with the maximum information and
maximum noise. On the other hand, for $\theta=\pi/2$ the measurement operator
have elements $(A_k)_{ij}=\delta_{ij}\frac{1}{\sqrt{d}}\sum_s
\delta_{k,j\oplus s}$ which lead to $F=1$ and $G=1/d$, {\em i.e.} to the
extreme case where the signal is preserved but the estimation has to be
random.  In fact, the linear combinations (\ref{omd}) are enough,
by varying the value of $\theta$, to explore the entire optimal trade-off
(\ref{Dbound}).
\section{Sequential measurements in a multiuser transmission line}
\label{s:Seq}
\begin{figure}[h]
\centerline{\includegraphics[width=0.8\textwidth]{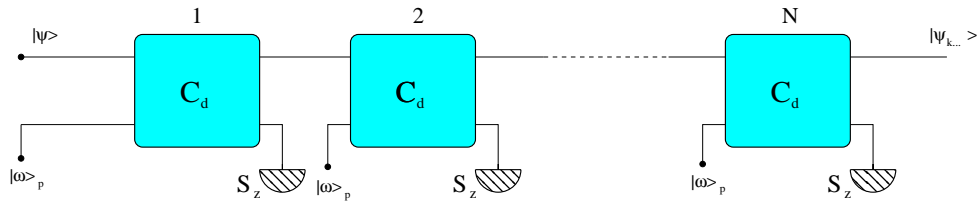}}
\caption{Sequential measurements in a multiuser transmission line.} 
\label{f:NScheme}
\end{figure}

Let us consider the multiuser transmission line schematically depicted 
in Fig. \ref{f:NScheme}. Each of the $N$ users detects the received 
signal by means of the device described in the previous section. At first
we consider all the N probes prepared in the same state ({\em i.e.}
 with the same value of $\theta$).
The measurements are sequential, {\em i.e} the $k$-th user measures the 
signal outgoing the $(k-1)$-th measurement stage. If the initial state
is $|\psi\rangle$ the first user obtains the result $k_1$ with 
probability $p_{k_1}=\langle \psi | A^\dag_{k_1} A_{k_1}|\psi\rangle$, 
leaving the conditional state $|\psi_{k_1}\rangle =
1/\sqrt{p_{k_1}} A_{k_1}|\psi\rangle$ to the subsequent user. The second
user obtains the result $k_2$ with conditional probability 
$p_{k_2,k_1}= \langle \psi | \left(A_{k_2} A_{k_1} \right)^\dag
A_{k_2}A_{k_1} |\psi\rangle$ 
and leave the conditional state 
$|\psi_{k_1,k_2}\rangle = \frac{1}{\sqrt{p_{k_2 k_1}}}
A_{k_2}A_{k_1}|\psi\rangle$. Analogously, the $N$-th user obtains 
the result $k_N$ with conditional probability 
\begin{align}
p_{k_N,k_{N-1},\dots,k_1} = \langle\psi|(A_{k_N}A_{k_{N-1}}\dots
A_{k_1})^{\dag} A_{k_N}A_{k_{N-1}}\dots A_{k_1}|\psi\rangle\:,
\end{align}
corresponding to the conditional state 
\begin{align}
|\psi_{k_N,k_{N-1},\dots, k_1}\rangle &= \frac{A_{k_N}A_{k_{N-1}}\dots
A_{k_1}|\psi\rangle}{\sqrt{p_{k_N,k_{N-1},\dots,k_1}}}\:.
\end{align}
\par\noindent
The unconditional probability of getting the outcome $k$ for the $N$-th user
is obtained upon summing over all possible results obtained in the previous
$N-1$ steps, {\em i.e.}
\begin{align}
p_k &= \Tr\[ A_k \( \sum_{k_1,k_2,\dots,k_{N-1}}A_{k_{N-1}}A_{k_{N-2}}
\dots A_{k_1}|\psi\rangle\langle\psi|(A_{k_{N-1}}A_{k_{N-2}}\dots 
A_{k_1})^{\dag} \) A_k^{\dag} \] \label{prob}\:.
\end{align}
The corresponding conditional state reads as follows
\begin{align}
\varrho_{k}^N&= \frac{1}{p_k} A_k \( \sum_{k_1,k_2,\dots,k_{N-1}}
A_{k_{N-1}}A_{k_{N-2}}\dots A_{k_1}|\psi\rangle\langle\psi| 
(A_{k_{N-1}}A_{k_{N-2}}\dots A_{k_1})^{\dag} \) A_k^{\dag}\:.
\end{align}
Notice that by using the independence of the measurement steps, {\em i.e} 
$\[ A_{k_i},A_{k_j}\] = 0$, and the normalization condition $\sum_{k_i} 
A_{k_i}^{\dag} A_{k_i} = {\mathbbm I}$, $\forall i$, the unconditional 
probability can be written as $p_k =\Tr[| \psi\rangle\langle\psi| \:
A_k^{\dag}A_k]$, that is, $p_k$ does not depend on the number $N$ of measurement
steps. The above formula also indicates that the POVM describing the 
measurement of the $N$-th user is given by $\{ \Pi_k = A_k^{\dag} A_k\}$, 
which is the same POVM of the single-user scheme described in the previous 
section. Using the {\em single-measure inference rule} $k \rightarrow
|k\rangle$, with $|k\rangle$ eigenstate of $S_z$, the estimation fidelity 
$G$ does not depend on the number of measurement steps, and is equal to 
the single-user one of Eq. (\ref{DfidsG}).
\par
The transmission fidelity at the $N$-th step for a given input signal 
$|\psi\rangle$ is given by 
\begin{align}
F_{\psi,N} &= \sum_{k} p_{k} \Tr[|\psi\rangle\langle\psi| 
\varrho_{k}^{N}] = \sum_{k,k_1,k_2,\dots,k_{N-1}} 
|\langle\psi|A_kA_{k_{N-1}}A_{k_{N-2}}\dots A_{k_1}|\psi\rangle|^2 
= \sum_{\{k_i\}} |\langle\psi|A_{\{k_i\}}|\psi\rangle|^2 \nonumber 
\end{align}
where
$ A_{\{k_i\}} = A_k A_{k_{N-1}}A_{k_{N-2}} \dots A_{k_1} $
and
$ \sum_{\{k_i\}} = \sum_k \sum_{k_{N-1}} \sum_{k_{N-2}} \dots \sum_{k_1} $ 
denotes sum over all indices $k_i$. 
Since $\sum_{\{k_i\}} A_{\{k_i\}}^{\dag}A_{\{k_i\}} = {\mathbbm I}$, we can 
evaluate the average fidelity by means of (\ref{fidseF}) with the substitution 
$k \rightarrow \{k_i\}$.
Since all the operators commute, the sum can be evaluated as follows 
\begin{align}
\sum_{\{k_i\}} \left| \hbox{Tr} \left[A_{\{k_i\}}\right] \right|^2 & = 
\sum_{n_0}^N\sum_{n_1}^{N-n_0} \dots \sum_{n_{d-2}}^{N-(n_0 + \dots + n_{d-3})} 
\frac{N!}{n_0!n_1!\dots n_{d-2}!(N-(n_0+n_1+\dots n_{d-2}))!} \nonumber \\
& \times \left| \Tr\left[A_0^{n_0} A_1^{n_1} \dots A_{d-2}^{n_{d-2}} 
A_{d-1}^{N - (n_0 + \dots + n_{d-2})}\right] \right|^2 \nonumber 
\end{align}
where the matrix elements of the operator under trace are given by 
\begin{align}
\left(A_0^{n_0} A_1^{n_1} \dots A_{d-2}^{n_{d-2}} 
A_{d-1}^{N - (n_0 + \dots + n_{d-2})} \right)_{ij} = 
\delta_{ij}J(\theta)^{N-n_i} L(\theta)^{n_i} 
\end{align}
with $L(\theta) = \left[ (\gamma \sin\theta)/\sqrt{d} + 
\cos\theta \right]$, $J(\theta) =\left[ (\gamma 
\sin\theta)/\sqrt{d}\right]$ and $n_{d-1}=N-(n_0 + 
\dots + n_{d-2})$. 
For qubit ($d=2$) and qutrit ($d=3$) we obtain
\begin{align}
F_N(d=2) &= \frac{1}{3} \left[ 2 + \sin^{2N}\theta\right]\qquad 
F_N(d=3)= \frac{1}{2} \left[ 1 + \sin^{2N}\theta\right]\:. 
\end{align}
\begin{figure}[h]
\centerline{\includegraphics[width=0.8\textwidth]{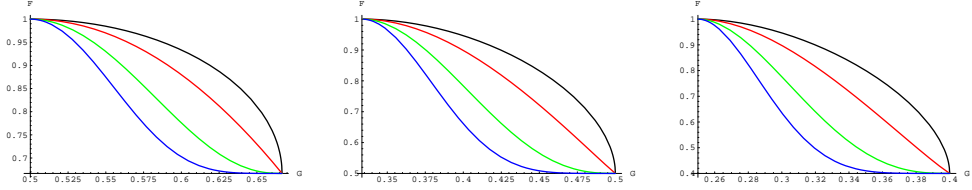}}
\caption{Information/disturbance trade-off for qudit in a $N$-users 
transmission line. (Black): $N=1$; (\textcolor{red}{Red}): $N=2$; 
(\textcolor{green}{Green}): $N=5$; (\textcolor{blue}{Blue}): 
$N=10$. From left to right the trade-off for $d = \{ 2, 3, 4\}$.}
\label{f:d234}
\end{figure}
\par\noindent
The corresponding information/disturbance trade-offs 
$F_N=F_N(G)$ for dimension  $d=\{2,3,4\}$ are depicted in Fig.
\ref{f:d234} for different values of $N$. As it is apparent from the plots, 
the trade-off degrades with the number of users $N$.  
\par
Let us now consider a different estimation strategy for the scheme depicted in 
Fig. \ref{f:NScheme}: where the $N$-th user has at disposal the
whole set of outcomes $\{k_i\} = \{ k_1, k_2, \dots, k_N\}$. The dynamics of 
the scheme is described by the overall measurement operators $A_{\{k_i\}}$.
The transmission fidelity does not change whereas the estimation fidelity
should be evaluated taking into account the global information coming from the 
whole set of outcomes. The {\em collective inference rule} is given by  
$\{k_i\} \rightarrow \varrho_{\{k_i\}}$ 
where 
\begin{align} 
\varrho_{\{k_i\}} = \frac{n_0}{N}|0\rangle\langle 0| + \frac{n_1}{N}
|1\rangle\langle 1| + \dots + \frac{N - (n_0 + n_1 + \dots + n_{d-2})}{N}
|d-1\rangle\langle d-1| 
\end{align}
and $n_j$ is the number of outcomes $j = \{ 0, 1,\dots, d-1 \}$ occurring 
in the sequence ${\{k_i\}}$. 
The explicit evaluation of the estimation fidelity
according to this rule gives the same results of Eq. (\ref{DfidsG})
for dimensions $d = \{2, 3, 4\}$ (we have been not not able to prove 
this for any value of $d$). In turn, this means that the trade-off
is not altered and shows that taking account collectively the whole set of 
outcomes does not lead to better performances.
\par
As it is apparent from Fig. \ref{f:d234} there is a region in the $(G,F)$ plane 
between the optimal trade-off ($N=1$) and the curve corresponding to $N=2$. 
A question arises on whether the $N=2$ scheme may be optimized in order to
reach points in this region. The answer is affirmative, by using a suitable 
preparation of the probe qudits. In the following we explicitly show how the
optimization procedure works in the case of qubit.
\par
\begin{figure}[h]
\centerline{\includegraphics[width=0.4\textwidth]{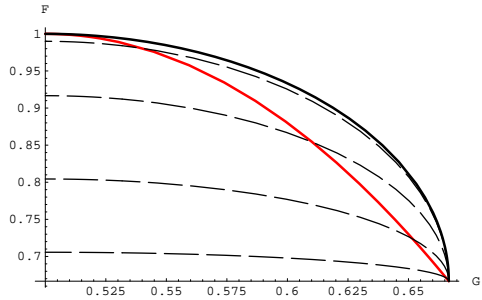}}
\caption{Information/disturbance trade-off for qubits in a $2$-user line with
different probes' parameter $\theta_A$ and $\theta_B$. In
\textcolor{red}{red} the trade-off obtained for $\theta_B=\theta_A$ and
varying this single parameter. The black solid line corresponds to 
the optimal trade-off obtained 
for $\theta_A=\pi/2$ and varying $\theta_B$. Black dashed curves correspond to
trade-offs $F=F_{2,\theta_A}(G)$ obtained varying $\theta_B$ at fixed values
of $\theta_A$. From top to bottom $\theta_A=\{4\pi/9 ; \pi/3; 2\pi/9 ; \pi/9
\}$.} \label{f:SeqRepAB} 
\end{figure}

 
Consider two users $A$ and $B$ which prepare their probes with different
parameter $\theta_A$ and $\theta_B$ and then perform a sequential measurement
using each the optimal scheme of the previous section. After the second step, 
the estimation fidelity $G$ is again the optimal one given by Eq. (\ref{DfidsG}) 
and depends only on the parameter $\theta_B$.  On the other hand, the transmission 
fidelity can be calculated by means of (\ref{fidseF}): the measurement 
operators $A_{\{k_i\}}= A_{k_B}A_{k_A}$ are given by 
\begin{align}
A_0 A_0 &=
\left(
\begin{array}{cc}
 L(\theta_A) L(\theta_B) & 0 \\
0 & J(\theta_A) J(\theta_B) 
\end{array}
\right)
\qquad 
A_0 A_1 =
\left(
\begin{array}{cc}
L(\theta_B) J(\theta_A) & 0 \\
0 &L(\theta_A) J(\theta_B)
\end{array} 
\right)
\nonumber \\
A_1 A_0 &=
\left( 
\begin{array}{cc}
L(\theta_A) J(\theta_B) & 0 \\ 
0 & L(\theta_B) J(\theta_A) 
\end{array}
\right)
\qquad
A_1 A_1 = 
\left(
\begin{array}{cc}
J(\theta_A) J(\theta_B) & 0 \\
0 & L(\theta_A) L(\theta_B) 
\end{array}
\right)
\end{align}
The fidelity is given by
\begin{align}
F_2(\theta_A,\theta_B) &= \frac{1}{24} \left[ 18 - 2 \cos(2\theta_A) - 
2\cos(2\theta_B) + \cos(2(\theta_A-\theta_B)) + \cos(2(\theta_A - 
\theta_B)) \right]\:.
\end{align}
The corresponding trade-off is shown in Fig. \ref{f:SeqRepAB}. As a matter of 
fact, the second user can tune the probe parameter $\theta_B$ to achieve points 
between the optimal trade-off and the curve obtained for $\theta_A=\theta_B$.
In particular, if $\theta_A=\pi/2$, {\em i.e.} if the first scheme is a {\em
blind} repeater, then the optimal trade-off can be re-obtained by varying
$\theta_B$. In the general case, the estimation fidelity can take all the 
allowed values, while the transmission fidelity $F_2$ take values in the 
range from $F_{min}=2/3$ to $F_2(\theta_A,\pi/2) = (5 - \cos 2\theta_A)/6$.  
In other words, the user $B$, by knowing the preparation of the first 
probe and varying the value of $\theta_B$, may achieve the desired
point on the curves $F=F_{2,\theta_A}(G)$ depicted in Fig.
\ref{f:SeqRepAB}, {\em i.e.} he can tune the trade-off and 
decide whether improving the  estimation fidelity or the transmission fidelity. 
Besides, this also means that by a suitable choice of both 
parameters $\theta_A$ and $\theta_B$, the entire region below the 
optimal trade-off (bound included) is accessible. 
The same argument may be applied to $d$-dimensional signals.
\section{Conclusions}\label{s:outro}
We have suggested a class of indirect measurement schemes, involving unitary
interactions of a signal qudit with a single probe qudit, which are suited to
extract information from a random set of qudit signals introducing the minimum
amount of disturbance. The schemes are indeed optimal, {\em i.e} correspond to 
estimation and transmission fidelities which saturate the ultimate bound
imposed by quantum mechanics.  The performances of optimal single-user schemes 
in extracting information by sequential measurements in a multiuser transmission 
line have been also investigated. We have explicitly evaluated fidelities and 
found that estimation  fidelity does not depend on the number of users, neither 
for single-measure inference nor for collective one, whereas the transmission fidelity 
decreases with the number of steps. The resulting trade-off is no longer optimal 
and degrades with increasing $N$. Optimality can be restored by a suitable 
preparation of the probe states: the optimization procedure have been
explicitly reported for qubit 2-user case.

\ack
This work has been supported by MIUR through the project PRIN-2005024254-002.


\end{document}